\documentclass[%
  conference,%
]{IEEEtran}

\usepackage{cite}
\usepackage{amsmath,amssymb,amsfonts}
\allowdisplaybreaks

\usepackage{booktabs}
\usepackage{url}

\usepackage{paralist}

\usepackage[UKenglish]{babel}

\usepackage{graphicx}
\graphicspath{{./images/}}
\usepackage{placeins}
\usepackage{subcaption}
\usepackage{enumitem}
\usepackage{balance}
\usepackage{booktabs}  %
\usepackage{colortbl}
\usepackage{flushend}

\usepackage[%
  group-minimum-digits=4,%
  list-final-separator={, and },%
  add-integer-zero=false,%
  free-standing-units,%
  binary-units,%
  detect-weight=true,%
  detect-inline-weight=math,%
]{siunitx}

\usepackage{microtype}

\usepackage{hyperref}

\usepackage[
colorinlistoftodos,prependcaption,textsize=tiny
]{todonotes}

\usepackage{xspace}

\usepackage[capitalise]{cleveref}

\crefformat{footnote}{#2\footnotemark[#1]#3}

\usepackage{tabularx}

\newcommand{\summary}[2]{
    \vspace{1em}
    \noindent
    \colorbox{gray!20}{%
        \parbox{.97\linewidth}{%
                \textbf{Summary \textit{(#1)}}
            #2
        }%
    }%
}%

\usepackage{csquotes}
\usepackage{csvsimple}
\usepackage{listings}
\usepackage{xcolor}
\usepackage{color,soul}

\definecolor{codegreen}{rgb}{0,0.6,0}
\definecolor{codegray}{rgb}{0.5,0.5,0.5}
\definecolor{codepurple}{rgb}{0.58,0,0.82}
\definecolor{backcolour}{rgb}{0.95,0.95,0.92}

\lstdefinestyle{mystyle}{
  backgroundcolor=\color{backcolour},
  commentstyle=\color{codegreen},
  keywordstyle=\color{magenta},
  numberstyle=\tiny\color{codegray},
  stringstyle=\color{codepurple},
  basicstyle=\ttfamily\footnotesize,
  breakatwhitespace=false,
  breaklines=true,
  breakindent=10pt,
  captionpos=b,
  keepspaces=true,
  numbers=left,
  numbersep=5pt,
  showspaces=false,
  showstringspaces=false,
  showtabs=false,
  tabsize=2
}

\makeatletter
\newcommand{\linebreakand}{%
  \end{@IEEEauthorhalign}
  \hfill\mbox{}\par
  \mbox{}\hfill\begin{@IEEEauthorhalign}
}
\makeatother

\newcommand{\baselinePrecision}{\SI{60.3}{\percent}\xspace}
\newcommand{\baselineRecall}{\SI{76.0}{\percent}\xspace}

\newcommand{\baselineFOne}{\SI{67.3}{\percent}\xspace}

\newcommand{\entropyDtOnePrecisionMeanPercent}{\SI{70.5}{\percent}\xspace}

\newcommand{\entropyDtOneRecallMeanPercent}{\SI{61.0}{\percent}\xspace}

\newcommand{\entropyDtOneFOneMeanPercent}{\SI{65.3}{\percent}\xspace}

\newcommand{\entropyDtOneThresholdRelativeStdPercent}{\SI{5.1}{\percent}\xspace}

\newcommand{\flipRateConstantDtOnePrecisionMeanPercent}{\SI{68.1}{\percent}\xspace}

\newcommand{\flipRateConstantDtOneRecallMeanPercent}{\SI{58.0}{\percent}\xspace}

\newcommand{\flipRateConstantDtOneFOneMeanPercent}{\SI{62.5}{\percent}\xspace}

\newcommand{\flipRateConstantDtOneThresholdRelativeStdPercent}{\SI{21.3}{\percent}\xspace}

\newcommand{\flipRateLinearDtOnePrecisionMeanPercent}{\SI{74.8}{\percent}\xspace}

\newcommand{\flipRateLinearDtOneRecallMeanPercent}{\SI{56.0}{\percent}\xspace}

\newcommand{\flipRateLinearDtOneFOneMeanPercent}{\SI{63.9}{\percent}\xspace}

\newcommand{\flipRateLinearDtOneThresholdRelativeStdPercent}{\SI{9.3}{\percent}\xspace}

\newcommand{\flipRateExponentialDtOnePrecisionMeanPercent}{\SI{73.6}{\percent}\xspace}

\newcommand{\flipRateExponentialDtOneRecallMeanPercent}{\SI{55.0}{\percent}\xspace}

\newcommand{\flipRateExponentialDtOneFOneMeanPercent}{\SI{62.7}{\percent}\xspace}

\newcommand{\flipRateExponentialDtOneThresholdRelativeStdPercent}{\SI{15.7}{\percent}\xspace}

\newcommand{\flipRateReciprocalDtOnePrecisionMeanPercent}{\SI{76.3}{\percent}\xspace}

\newcommand{\flipRateReciprocalDtOneRecallMeanPercent}{\SI{58.0}{\percent}\xspace}

\newcommand{\flipRateReciprocalDtOneFOneMeanPercent}{\SI{65.8}{\percent}\xspace}

\newcommand{\flipRateReciprocalDtOneThresholdRelativeStdPercent}{\SI{6.7}{\percent}\xspace}

\newcommand{\flipRateReciprocalSquaredDtOnePrecisionMeanPercent}{\SI{79.1}{\percent}\xspace}

\newcommand{\flipRateReciprocalSquaredDtOneRecallMeanPercent}{\SI{61.0}{\percent}\xspace}

\newcommand{\flipRateReciprocalSquaredDtOneFOneMeanPercent}{\SI{68.8}{\percent}\xspace}

\newcommand{\flipRateReciprocalSquaredDtOneThresholdRelativeStdPercent}{\SI{5.5}{\percent}\xspace}

\newcommand{\flipRateEwmaDtOnePrecisionMeanPercent}{\SI{79.3}{\percent}\xspace}

\newcommand{\flipRateEwmaDtOneRecallMeanPercent}{\SI{58.0}{\percent}\xspace}

\newcommand{\flipRateEwmaDtOneFOneMeanPercent}{\SI{66.9}{\percent}\xspace}

\newcommand{\flipRateEwmaDtOneThresholdRelativeStdPercent}{\SI{1.1}{\percent}\xspace}

\newcommand{\flipRateXmeanDurationGbPrecisionMeanPercent}{\SI{70.0}{\percent}\xspace}

\newcommand{\flipRateXmeanDurationGbRecallMeanPercent}{\SI{72.0}{\percent}\xspace}

\newcommand{\flipRateXmeanDurationGbFOneMeanPercent}{\SI{70.8}{\percent}\xspace}

\newcommand{\flipRateXmeanDurationDiffGbPrecisionMeanPercent}{\SI{82.6}{\percent}\xspace}

\newcommand{\flipRateXmeanDurationDiffGbRecallMeanPercent}{\SI{78.0}{\percent}\xspace}

\newcommand{\flipRateXmeanDurationDiffGbFOneMeanPercent}{\SI{79.6}{\percent}\xspace}

\newcommand{\flipRateXmeanDurationXmeanDurationDiffGbPrecisionMeanPercent}{\SI{81.0}{\percent}\xspace}

\newcommand{\flipRateXmeanDurationXmeanDurationDiffGbRecallMeanPercent}{\SI{84.0}{\percent}\xspace}

\newcommand{\flipRateXmeanDurationXmeanDurationDiffGbFOneMeanPercent}{\SI{82.4}{\percent}\xspace}

\newcommand{\flipRateXcodeChurnGbPrecisionMeanPercent}{\SI{93.1}{\percent}\xspace}

\newcommand{\flipRateXcodeChurnGbRecallMeanPercent}{\SI{92.0}{\percent}\xspace}

\newcommand{\flipRateXcodeChurnGbFOneMeanPercent}{\SI{92.4}{\percent}\xspace}

\newcommand{\flipRateXcppFiftyfourXchangedFilesGbPrecisionMeanPercent}{\SI{95.2}{\percent}\xspace}

\newcommand{\flipRateXcppFiftyfourXchangedFilesGbRecallMeanPercent}{\SI{96.0}{\percent}\xspace}

\newcommand{\flipRateXcppFiftyfourXchangedFilesGbFOneMeanPercent}{\SI{95.5}{\percent}\xspace}

\newcommand{\fullGbPrecisionMeanPercent}{\SI{91.1}{\percent}\xspace}

\newcommand{\fullGbRecallMeanPercent}{\SI{90.0}{\percent}\xspace}

\newcommand{\fullGbFOneMeanPercent}{\SI{90.4}{\percent}\xspace}

\newcommand\copyrighttext{%
  \footnotesize
Accepted at ICST 2023.\@ \copyright~2023 IEEE.
Personal use of this material is permitted.
Permission from IEEE must be obtained for all other uses, in any current or future media, including
reprinting/republishing this material for advertising or promotional purposes, creating new collective works,
for resale or redistribution to servers or lists, or reuse of any copyrighted component of this work in other works.
}
\newcommand\copyrightnotice{%
    \begin{tikzpicture}[remember picture,overlay]
        \node[anchor=south,yshift=10pt] at (current page.south) {\fbox{\parbox{\dimexpr\textwidth-\fboxsep-\fboxrule\relax}{\copyrighttext}}};
    \end{tikzpicture}%
}

\begin{document}

\title{Practical Flaky Test Prediction using Common Code~Evolution and Test History Data}

\author{%
 \IEEEauthorblockN{Martin Gruber$^{\ddag*}$}%
 \IEEEauthorblockA{%
  martin.gr.gruber@bmw.de%
 }%
 \and%
 \IEEEauthorblockN{
    Michael Heine$^{\S*}$,
    Norbert Oster$^\S$,
    Michael Philippsen$^\S$
 }%
 \IEEEauthorblockA{%
  \{michael.heine $|$ norbert.oster $|$ michael.philippsen\}@fau.de%
 }%
 \and%
 \IEEEauthorblockN{Gordon Fraser$^\ddag$}%
 \IEEEauthorblockA{%
  gordon.fraser@uni-passau.de%
 }%
 \linebreakand%
 \vspace{-15pt}%
 \IEEEauthorblockN{}%
 \IEEEauthorblockA{
    $^{\S}$Programming Systems Group, Friedrich-Alexander-Universität Erlangen-Nürnberg (FAU), Germany\\
    $^\ddag$University of Passau, Germany \hspace{20pt}
    $^*$BMW Group, Munich, Germany
 }%
}

\maketitle

\begin{abstract}

Non-deterministically behaving test cases cause developers to lose trust in their regression test suites and to eventually ignore failures.
Detecting flaky tests is therefore a crucial task in maintaining code quality, as it builds the necessary foundation for any form of systematic response to flakiness, such as test quarantining or automated debugging.
Previous research has proposed various methods to detect flakiness, but when trying to deploy these in an industrial context, their reliance on instrumentation, test reruns, or language-specific~artifacts was inhibitive.
In this paper, we therefore investigate the prediction of flaky tests without such requirements on the underlying programming language, CI, build or test execution framework.
Instead, we rely only on the most commonly available artifacts, namely the tests' outcomes and durations, as well as basic information about the code evolution to build predictive models capable of detecting flakiness.
Furthermore, our approach does not require additional reruns, since it gathers this data from existing test executions.
We trained several established classifiers on the suggested features and evaluated their performance on a large-scale industrial software system, from which we collected a data set of 100 flaky and 100 non-flaky test- and code-histories.
The best model was able to achieve an F1-score of \flipRateXcppFiftyfourXchangedFilesGbFOneMeanPercent using only 3 features: the tests' flip rates, the number of changes to source files in the last 54 days, as well as the number of changed files in the most recent pull request.

\end{abstract}

\copyrightnotice

\section{Introduction}

As the large corpus of both academic literature and online blog articles shows, the issue of test flakiness is regarded as a major concern in the software testing community~\cite{parry2021survey, fowler2011eradicating, micco2016flaky}.
This discussion has brought forth several approaches aiming to mitigate the problem by
visualizing the behavior of flaky tests~\cite{oezal2021how, champier2017hunting, palmer2019test},
quantifying their impact~\cite{raine2020reducing}, or quarantining them~\cite{lam2020study, fowler2011eradicating}
as well as by
automatically debugging~\cite{lam2019root, terragni2020container, celalziftci2020de, akli2022predicting}, reducing~\cite{lam2020study},
or even fixing~\cite{shi2019ifixflakies, wang2022ipflakies} flakiness.
What they have in common is that all of these methods rely on the availability of effective techniques to accurately detect which tests are flaky and should therefore be displayed on a dashboard, trigger an alarm, or be intensively analyzed.
While determining if a given test failure is caused by flakiness might not be a complicated task for a skilled developer, manual inspection becomes infeasible for larger projects that make hundreds of changes per day and therefore run thousands of regression tests.
Hence, there is a need for detecting flaky tests automatically.

To address this demand, researchers have proposed many flakiness detection approaches~\cite{
    alshammari2021flakeflagger, %
    parry2022evaluating, %
    shi2016detecting, %
    bell2018deflaker, %
    dutta2020detecting, %
    silva2020shake, %
    lam2019idflakies, %
    kowalczyk2020modeling, %
    machalica2020how, %
    king2018towards,
    fatima2022flakify,
    qin2022peeler, %
    pinto2020what,
    verdecchia2021know   
}.
We found most of these difficult to implement in an industrial setting, since they rely on code instrumentation~\cite{
    alshammari2021flakeflagger, %
    parry2022evaluating, %
    dutta2020detecting, %
    shi2016detecting, %
    silva2020shake, %
    bell2018deflaker %
},
which is not always implementable,
demand multiple test reruns~\cite{
    lam2019idflakies, %
    dutta2020detecting, %
    silva2020shake, %
    kowalczyk2020modeling, %
    machalica2020how %
},
which is computationally demanding,
or require language-specific artifacts~\cite{
    qin2022peeler, %
    fatima2022flakify,
    pinto2020what,
    verdecchia2021know,
    king2018towards
}.
Our observation is underlined by a recent study that found the adoption rates of automated flakiness detection techniques among practitioners to be poor~\cite{gruber2022survey}.

Despite the limitations of existing approaches, there exists information that is both widely and easily available in most industrial contexts, and that we hypothesize to be sufficient in order to build predictive models to detect test flakiness.
In particular, this data includes the tests' outcomes and durations from existing test executions (the test history), as well as code churn information, such as the number of changed files in the current pull request (the evolution of code).
Using a large-scale industrial software project as subject for evaluation, we aim to assess if this information can indeed be applied for effectively detecting flaky tests.
To do so, we sample test data from the case study software project, and build and evaluate different feature combinations and multiple established binary classifiers.

In detail, our key contributions are:
\begin{itemize}
    \item a compilation and implementation of widely and easily available candidate features for flakiness detection,
    \item a data set of 100 flaky and 100 non-flaky tests derived from a real industrial software product, and
    \item a detailed evaluation of the flakiness detection capabilities of this information on the data set.
\end{itemize}

Our results suggest that the evolution of code and the history of tests are powerful predictors for test flakiness, with the best performing model achieving an F1-score of \flipRateXcppFiftyfourXchangedFilesGbFOneMeanPercent while using only 3 features:
the weighted flip rate, the number of source code changes in the last 54 days, and the number of changed files in the current pull request.
When comparing different weighting algorithms for the flip rate, we found that stronger decaying functions yield better results.
By examining different feature combinations, we found that code churn information plays a vital role for our models: without it, the predictive performance dropped to an F1-score of \flipRateXmeanDurationXmeanDurationDiffGbFOneMeanPercent.
Overall, our study demonstrates that common code evolution and test history data can be effectively applied to detect flakiness and therefore offer a compelling alternative for projects that cannot meet the requirements of existing flakiness detection techniques.

\section{Background}%
\label{sec:background}

This section discusses existing flakiness detection approaches proposed by researchers, grouped by the main requirements they insist on.
Furthermore, we look at techniques developed by practitioners and discuss their demands.

\subsection{Using Instrumentation or Language-specific Artifacts to Detect Flakiness}%
\label{sec:background_instrumentation}

Many existing flakiness detection approaches rely on information extracted at runtime:
FlakeFlagger~\cite{alshammari2021flakeflagger} and Flake16~\cite{parry2022evaluating} measure properties such as API usage, file-system access, memory usage, and threading behavior to extract features for training binary classifiers to distinguish flaky from non-flaky tests.
Others go a step further and mutate the execution environment to expose flakiness by setting seeds of random number generators~\cite{dutta2020detecting}, switching implementations of methods with non-deterministic specifications~\cite{shi2016detecting}, or adding noise to the execution environment~\cite{silva2020shake}.
Another prominent example is DeFlaker~\cite{bell2018deflaker}, which---given an existing CI history of a test case (test outcomes and changes between them)---performs a differential coverage analysis to determine if a change could have affected a test's outcome.
If this was not the case and the test outcome still changed, the test failure was not caused by the change and is therefore a flaky failure.

While these methods have been shown to be effective, the target audience able to use them is potentially limited, since the required instrumentation cannot always be implemented.
For example, even gathering code coverage, which is a mostly straightforward and tool-wise well-supported undertaking, can be infeasible under certain circumstances, such as in resource-constrained embedded systems~\cite{elbaum2014techniques}.
Furthermore, even if the required information could theoretically be gathered, the implementations of these approaches are specific to certain programming languages or test execution frameworks, e.g., Maven~\cite{bell2018deflaker, alshammari2021flakeflagger}, and transferring them to other environments might come with a considerable ramp-up effort.
Even static methods that do not rely on instrumentation mention this as a threat to validity:
\enquote*{While our design is generic and does not present specific
programming-language constraints, our implementation was
guided by the availability of artefacts (e.g., program parser)
and benchmarks for evaluation.}~\cite{qin2022peeler}.
This shows that the availability of language-specific artifacts
is a concern not only to dynamic, but also to static and hybrid approaches, that use them to extract code smells~\cite{fatima2022flakify}, 
data dependencies~\cite{qin2022peeler}, or
the vocabulary of tests~\cite{pinto2020what, verdecchia2021know},
or to measure complexity metrics~\cite{king2018towards}.

\subsection{Using Reruns to Detect Flakiness}%
\label{sec:background_reruns}

Repeatedly executing tests on the same version of the system under test is the most commonly used approach to detect flaky tests~\cite{gruber2022survey} and is being systematically applied by large software companies such as Microsoft~\cite{lam2020study} and Google~\cite{micco2017state}.
While being tool-wise well-supported through plugins to popular test execution frameworks such as Maven~\cite{mavenRerunPlugin} and PyTest~\cite{pythonRerunPlugin}, the computational effort related to this technique can be immense:
It has been shown that on average at least 100 reruns are required to detect just half of all flaky tests within a test suite~\cite{gruber2021empirical, alshammari2021flakeflagger}.
This is particularly infeasible for larger tests, such as system tests or hardware-in-the-loop tests, where flakiness has been shown to be significantly more common than in smaller tests~\cite{listfield2017where}.
The unsustainability of rerunning tests to detect flakiness is also reflected in the fact that even some test execution frameworks that do offer rerun-on-failure options discourage their usage~\cite{bazelAttributeFlaky}.

To minimize the number of reruns needed to expose flakiness, researchers have proposed several approaches:
iDFlakies~\cite{lam2019idflakies} reruns tests first in random, then in specific orders, and is therefore also capable of distinguishing order-dependent from non-order-dependent flaky tests.
Flash~\cite{dutta2020detecting} and Shaker~\cite{silva2020shake} mutate the execution environment to expose flakiness.
While these techniques undoubtedly outperform the normal rerun strategy, they still heavily rely on repeatedly executing tests.
This can be infeasible in some domains:
\enquote*{In many industrial contexts [...] it is not useful to look for flaky tests, due to frequent changes made in the underlying SW, HW and TW [testware] in a CI development paradigm.}~\cite{strandberg2020intermittently}.

Given historic test executions, Kowalczyk et al.~\cite{kowalczyk2020modeling} classified tests as flaky by calculating their outcomes' entropy and flip rate.
While not conducting additional reruns, they assume that in the past the tests have been executed multiple times on the same software version, which they call an \textit{epoch}, which might be unrealistic:
\enquote*{Since the perceived business value of retesting is limited when nothing has changed, we cannot expect to see much repeated testing on unchanged SW, TW and HW.}~\cite{strandberg2020intermittently}.
Nevertheless, flip rate and entropy can also be calculated without this assumption.
We quantitatively evaluate this in~\cref{sec:evaluation_rq1}.

\subsection{Flakiness Detection in Practice}%
\label{sec:background_flakiness_detection_in_practice}

To spot test flakiness, practitioners~\cite{
    palmer2019test, %
    liviu2019machine, %
    machalica2020how, %
    champier2019continuous %
} often create dashboards to visualize test outcomes over time or list flaky tests, and even commercial tools exist that offer this functionality~\cite{
    gradle2020best, %
    oezal2021how, %
    launchableFlakyTestInsights, %
    rudolphbuildpulse %
}.
In order to appeal to a wide and diverse audience, most of them avoid making assumptions about the underlying tests and refrain from triggering or intercepting test executions.
However, many of them also rely on existing reruns to distinguish flaky from broken tests, which might be infeasible, as discussed in~\cref{sec:background_reruns}.

\begin{figure*}
    \centering
    \fbox{\includegraphics[width=\textwidth]{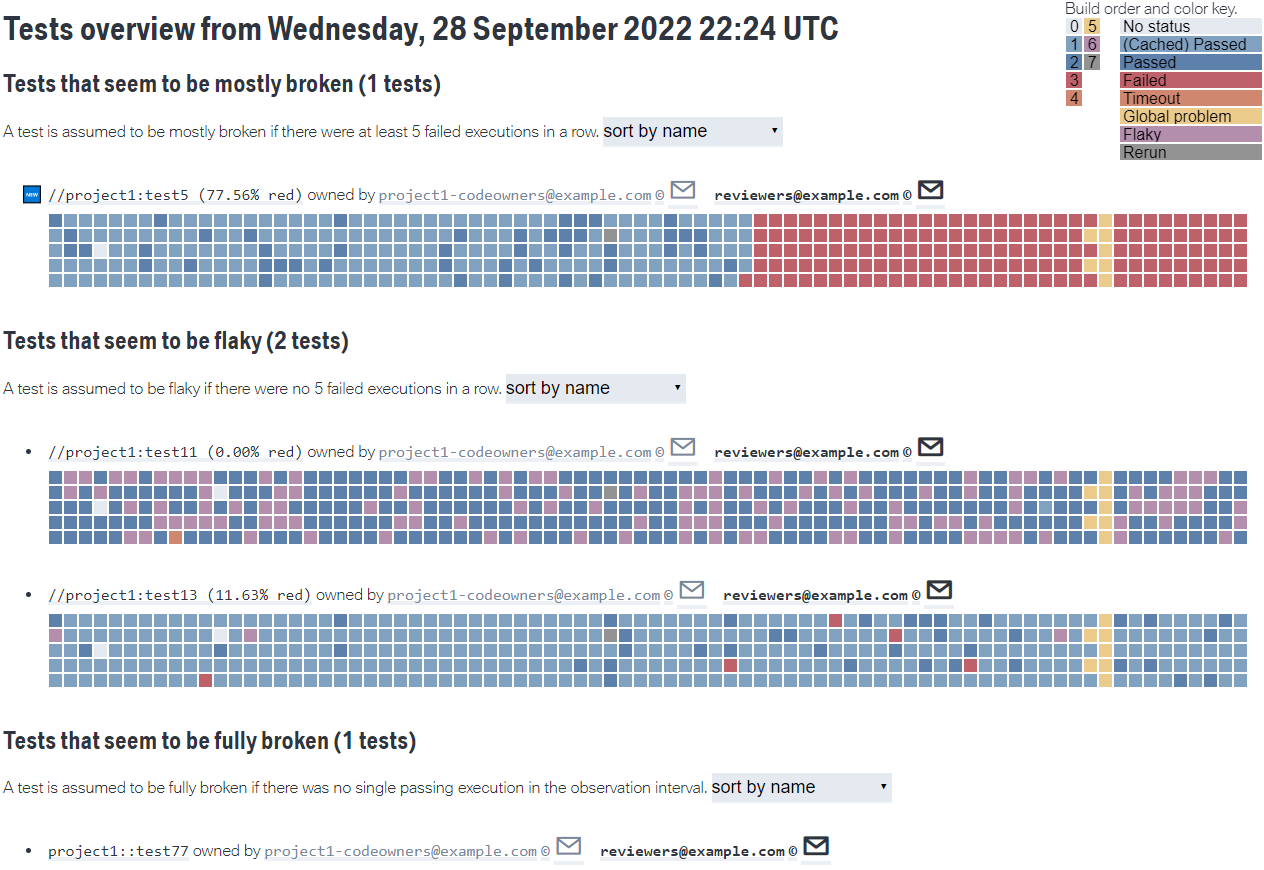}}
    \caption{CTRA (CI Test Results Analysis) frontend. Timelines are ordered column-wise, left to right.}
    \label{fig:ctra_screenshot}
\end{figure*}

\cref{fig:ctra_screenshot} shows a screenshot of CTRA (CI Test Results Analysis), an internal tool developed by our industry partner, which is \enquote*{meant to support developers in monitoring CI tests with respect to test status and test flakiness}.
Among other features, it displays a timeline for each flaky and broken test case, depicting the results of its most recent executions, similar to the \textit{Odeneye} system developed by Spotify\cite{palmer2019test}.

To distinguish flaky from broken tests, it uses the following heuristics:
If a test case failed at least once within the last 400 builds, but never five times in a row, it is regarded as \textit{flaky}.
Alternatively, a test is also seen as flaky if a rerun on failure---which is configured for certain tests---passed, yielding test outcome \texttt{Flaky}.
If, on the other hand, a test case did fail at least five times in a row within that window but also exhibited passing test executions, it is regarded as \textit{mostly broken}.
Lastly, if a test showed only failing test executions, it is seen as \textit{fully broken}.
Tests that only passed are not shown on the dashboard.
The observation window size of 400 test executions was selected based on domain knowledge and should roughly reflect 24 hours of development.
Like other tools developed by practitioners, these heuristics make no assumptions about the underlying tests or their programming language, which is required since (1) CTRA is deployed on a multi-language project and has the aspiration to be usable for all tests in the project, and (2) CTRA is a strictly non-invasive visualization tool that should not trigger any test executions or make any code changes.
While CTRA uses information from existing reruns, our industry partner views reruns as a suboptimal strategy to detect flakiness, mainly due to the delayed test feedback.
For this reason, they are searching for more sustainable, scalable, and efficient ways to detect flakiness.
Their current approach will therefore serve as a baseline for our models.

Overall, we see a clear need for a generic, non-invasive, and computationally less demanding flakiness detection technique, coming from both our industry partner and other practitioners.

To the best of our knowledge, the only approach presented by researchers that meets these requirements is a technique presented by Strandberg et al.~\cite{strandberg2020intermittently}, who also conducted their study on an industrial system:
Using Markov chains, they effectively computed the unweighted flip rate.
However, their goal was not to accurately detect flaky tests, but to study intermittently failing tests, which are tests that showed different outcomes under a potentially changing environment (system under test, test code, hardware, ...), which therefore include both expected intermittent failures (i.e., regressions) and unexpected intermittent failures (i.e., flakiness).
In contrast, our goal is to distinguish between flaky failures and failures caused by regressions.
Building on their work, we extend and quantitatively evaluate the flakiness-detection capabilities of the flip rate, also in combination with other test history features, as well as code churn information.
The latter has also been suggested by Strandberg et al.~\cite{strandberg2020intermittently}.

\begin{figure*}
    \centering
    \includegraphics[width=\textwidth]{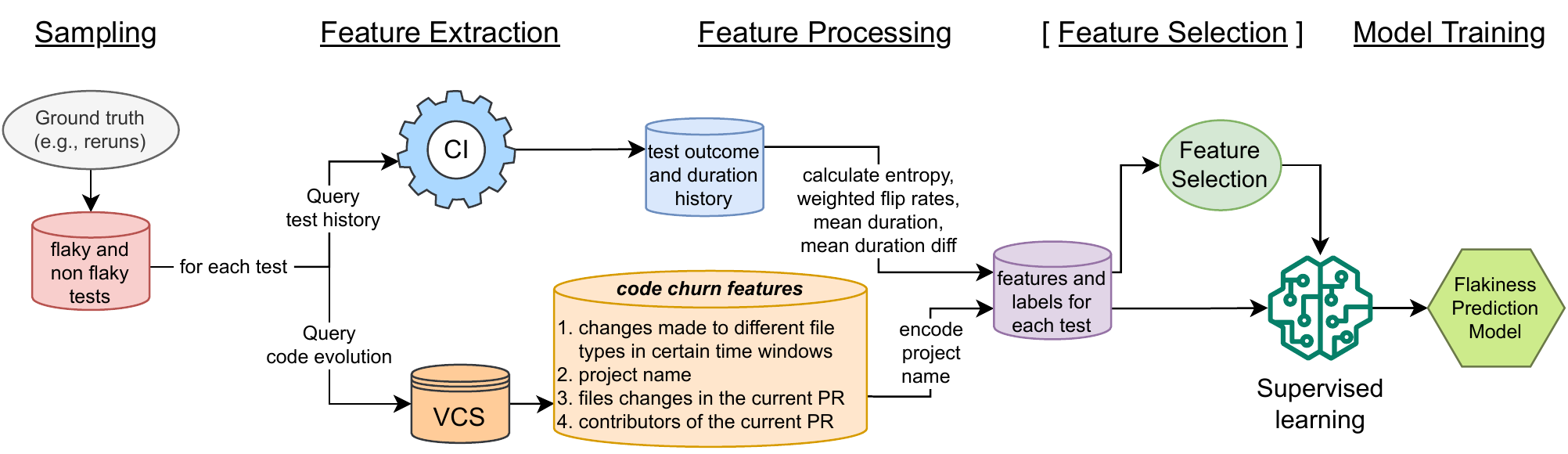}
    \caption{Overview of our approach.}
    \label{fig:approach_overview}
\end{figure*}

\section{Code Evolution and Test History-based Flakiness Detection}%
\label{sec:Methodology}

Having discussed existing flakiness detection approaches in~\cref{sec:background}, we see a mismatch between the strong technical demands of methods proposed by researchers, and the requirements of practitioners, asking for non-invasive, language-independent tools.
Therefore, our goal is to build a more generic technique for detecting flakiness that is easily accessible to a wide range of developers.
~\cref{fig:approach_overview} depicts an overview of our approach:
For each test case in a given training data set, we extract most commonly available information about existing test executions from the continuous integration system (test history), as well as code churn information from the version control system (code evolution).
Using this data,
together with ground truth labels coming for example from reruns,
we train a classifier to predict if a test with a given history is flaky.
The information we extract from the  continuous integration system are the test verdicts and durations from previous test executions.
We deem these suitable candidates as the former is the essential outcome of any form of testing and the latter is trivially measurable.
This claim is supported by the fact that both are part of JUnit XML~\cite{JUnitXMLSchema}, a common format for reporting test results that many test execution frameworks support and that is not limited to the Java ecosystem (e.g., PyTest~\cite{pytestJUnitXML}).
Below, we discuss our feature processing of test outcomes, test durations, and code churn information.

\subsection{Test Outcomes}

In a regression test scenario, developers experience flakiness in the form of sudden, unexpected test failures that likely disappear when executing the test case again on the same version or when testing the next changes.
Unlike regressions that cause a test to consistently fail until it is fixed, flakiness should therefore manifest itself in frequently altering test outcomes, or in other words, in a high flip rate.
This claim is supported by experts from Google reporting that about \SI{84}{\percent} of the transitions they observed from pass to fail involved a flaky test~\cite{micco2016flaky, micco2017state}.
The flip rate is formally defined as follows~\cite{kowalczyk2020modeling}:

The \textit{test result history}
$R$ of a test case $c_1$ is the list of outcomes that its last $n$ executions produced: $R(c_1) = (r_{1}, r_{2}, ..., r_{n})$. %
Each pair $(r_t, r_{t+1})$
in this list represents one transition.
The flip rate is the relative number of transitions in which the test outcome changed (i.e.,\ $r_t \neq r_{t+1}$):
\begin{equation*}
    flip\_rate(R) = 
        \sum_{t=1}^{n-1} \left(
        \frac{1}{n-1} \cdot \left\{
            \begin{array}{ll}
                1, & r_{t} \neq r_{t+1} \\
                0, & r_{t} = r_{t+1} \\
            \end{array}
            \right.
    \right)
\end{equation*}

Building on this definition, Kowalczyk et al.~\cite{kowalczyk2020modeling} introduced a weighted model that computes an exponentially weighted moving average, giving more weight to outputs of more recent executions.
Putting more emphasis on recent executions should allow the model to correctly label non-flaky tests that have been flaky in the past, but have recently been fixed.
These tests are not flaky anymore, but carry flaky behavior in their long-term history.
Since the optimal weighting depends on the development speed and the granularity of the changes, we extend their approach by applying and evaluating different decay functions, not only the exponential moving average.
The adjusted formula for calculating the flakiness score using a generic decay function $w$ is:
\begin{equation*}
    flip\_rate(R, w) = 
        \sum_{t=1}^{n-1} \left(
        \frac{w(t)}{\sum_{u=1}^{n-1} w(u)} \cdot \left\{
            \begin{array}{ll}
                1, & r_{t} \neq r_{t+1} \\
                0, & r_{t} = r_{t+1} \\
            \end{array}
            \right.
    \right)
\end{equation*}

\begin{figure}
    \centering
    \includegraphics[width=\linewidth]{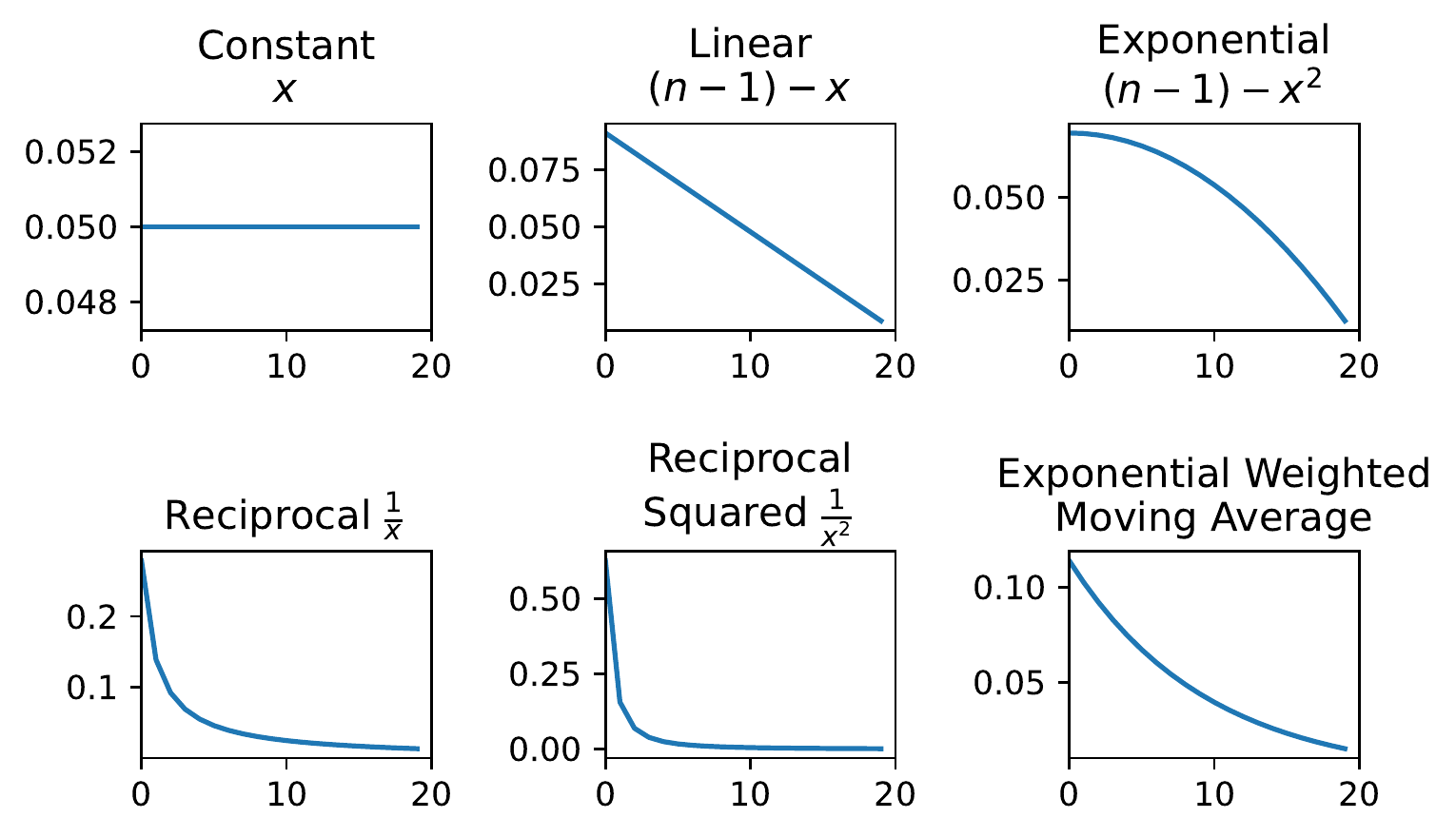}
    \caption{Decay functions for history length 21 (20 transitions).}
    \label{fig:decay_functions}
\end{figure}

As decay functions we use the constant\footnote{In this case, the formula is equal to the previous definition with $w(t) = 1$.} (i.e., unweighted), linear, exponential, reciprocal, and reciprocal-squared function (\cref{fig:decay_functions}). Furthermore, we include the exponential weighted moving average with a decay value of 0.1, which was empirically calibrated by Kowalczyk et al.~\cite{kowalczyk2020modeling}.

Apart from the flip rate, Kowalczyk et al.~\cite{kowalczyk2020modeling} also used the entropy of a test result history to detect flaky tests:
\begin{equation*}
    entropy(R) = -\sum_{i \in \{P, F\}} p(i) \textrm{ log } p(i)
\end{equation*}

with $p(i)$ being the probability of outcome $i$ (pass or fail).
Unlike the flip rate, the entropy of a test result history ignores the order in which the outcomes appear and only depends on the ratio of passes to failures.
We therefore suspect that the flip rate is going to be a better predictor for test flakiness.
Kowalczyk et al.~\cite{kowalczyk2020modeling} apply their weighting model by splitting test executions into chunks, calculating metrics (flip rate and entropy) for each chunk, and accumulating the chunk scores using an exponential moving average.
The splitting is done along epochs.
Within the same epoch, all test runs are conducted on the same version.
In contrast to them, we do not assume to have reruns on the same software versions, i.e., we have a chunk size of 1.
Under such conditions we cannot apply weighting to the entropy since the entropy of a chunk containing only one item is always 0.
But we can still apply a decay mechanism to the flip rate by weighting each transition.

\subsection{Test Duration}%
\label{sec:test_duration}

Apart from the verdict, the duration is another attribute of a test execution that is widely available, independently of programming languages, build-, CI-, and test-execution-frameworks.
The durations of multiple consecutive executions of the same test case form its \textit{test duration history}, which we process in two ways to coin features for classifying flaky tests.

The first feature derived from the test duration history is the mean test execution time (\textit{mean duration}).
Since larger tests have been shown to be significantly more flaky than smaller ones~\cite{listfield2017where}, we believe that this feature will help to distinguish flaky from non-flaky tests.
The second feature
is the difference between the mean duration of passing and failing runs (\textit{mean~duration~diff}).
Previous studies have shown that both concurrency and asynchronous waiting are major causes of flakiness~\cite{luo2014empirical, eck2019understanding, parry2022surveying, gruber2022survey}.
Since these aspects have significant impact on a test's duration through scheduling and timeouts, we suspect that they create characteristic patterns, by which we might be able to detect them.

In particular, we see two scenarios where flakiness might cause such a characteristic pattern in the duration difference between passing and failing test executions:
If a test queries a remote server that is not available or not reachable (e.g., due to authentication issues), the resulting flaky failure should appear sooner than a comparable non-flaky failure that occurs while processing the server's response.
This would lead to an above-average difference between passing and failing runs (fast failures).
On the other hand, a flaky failure might also be caused by timeouts while waiting for a response from a remote server.
In this case, the flaky failure takes longer than a comparable non-flaky failure, resulting in a below-average or even negative difference between passing and failing runs (slow failures).
To conclude, by calculating the difference between the mean passing and failing durations of a test case, we aim to effectively reverse engineer its dependence on external services from a black-box perspective, and we expect extreme values of this feature to hint towards flakiness.

\subsection{Code Churn Information}

Apart from the test execution history of a project, the evolution of its code offers another source for widely and easily available data.
Our intuition for how such information can be utilized to predict if a test is flaky is the following:
Provided with detailed information about changes to the system under test (code churn), a classifier should be able to learn that small changes or changes of certain types (for example to README files) do not have any impact on the code itself, so if we see different test outcomes under such changes (i.e., a high flip rate), the test is likely flaky.
In other words, we aim to exercise a similar reverse test selection process as DeFlaker~\cite{bell2018deflaker}, but without relying on data that requires instrumentation.
Utilizing code churn information to detect flakiness has also been suggested by previous research~\cite{strandberg2020intermittently, alshammari2021flakeflagger}.
Since we want our feature set to be independent from technical details of the underlying project, we refrain from looking at the content of the code changes, but only quantify and group them on the file level.
Specifically, we use the following list of features, which is inspired by Machalica et al.~\cite{machalica2019predictive} who proposed a predictive test selection approach:
\begin{enumerate}[label=(\alph*)]
    \item for each file extension, the number of changes made to such files in the last 3, 14, and 54 days
    (representing short-, medium-, and long-term changes),
    \item the project name (one-hot encoded),
    \item the number of modified files in the current pull request,
    \item the number of contributors of the current pull request.
\end{enumerate}

Note that depending on a concrete project, the size of the feature set (a) is the number of occurring file extensions times three
(due to the three time intervals of 3, 14, and 54~days).
For (b) there is a feature per project in the data set.
We discuss the concrete number of features that we extracted from the code evolution of our evaluation subject in \cref{sec:evaluation_dataset_featureCollection_codeEvolution}.
We implemented the extraction of these features by querying Git---the by far the most popular version control system~\cite{SOSurvey2022VCS}.

\subsection{Classifier Selection}%
\label{sec:methodology_classifier_selection}

We designed our approach to be agnostic to the applied classification algorithm, meaning that most supervised learning classifiers can be used to build models from our features.
For example, we encoded all features numerically, demanding no support for categorical or ordinal data from the classifier.
While it might be possible to achieve even better results using deep learning, we refrain from investigating this option, since these methods require large amounts of data, which may not be available in practice (for example, in our evaluation subject).
Instead, we focus on basic machine learning methods, aiming to create models that are easy to train and have limited complexity.

\section{Evaluation}

The test history and code churn features introduced in the previous section can be combined in several ways, resulting in different use cases.
Through our evaluation, we look at four different combinations, aiming to answer the following research questions:

\begin{description}
    \item[RQ1] How well can we detect flaky tests using only test outcomes?
    \item[RQ2] How well can we detect flaky tests using test outcomes and durations?
    \item[RQ3] How well can we detect flaky tests using test outcomes and code churn information?
    \item[RQ4] How well can we detect flaky tests using test outcomes, test durations, and code churn information?
\end{description}

Since we expect the test outcome to be always available, we include it in each feature set.
In RQ1 we do not add any further features, as this is exactly the information our baseline is using, which allows for a direct comparison of the feature processing methods.
To isolate the effect of additional features, we separately add the tests' durations (RQ2), and code churn information
(RQ3).
Finally, in RQ4, we consider the full feature set consisting of the tests' outcomes, their durations, as well as code churn information.

\subsection{Data Set}

\subsubsection{Evaluation subject}
Our evaluation subject is a large automotive software project.
It consists of nine repositories, organized as Git submodules, including the product software, as well as code for the hardware platform and the CI system.
The main programming languages are C and C++ with a combined total of more than 10 million lines of code spread across more than \num{60000} source files (measured via CLOC~\cite{adanial_cloc}, generated code was excluded).
Other languages include, for example, Python ($\sim$\num{800000} LoC) and Bash ($\sim$\num{50000} LoC).
It contains more than %
\num{45000} tests, ranging from simple unit tests to complex simulations.
The tests are mostly written in C++ and Python, however, Python tests are not necessarily used to test Python code, but oftentimes as wrappers for C code.%

\subsubsection{Testing strategy}
All tests are built and executed using \textit{Bazel}\cite{bazel}, which among other features provides support for many languages.
To schedule test executions, the project uses a CI system that is organized in several pipelines along the pull request-driven development model.
The \enquote{check} pipeline runs whenever a commit is pushed to a pull request branch.
It consists of linting and build steps for different compilers and other configurations, as well as the execution of---depending on the configuration---up to about \num{15000} tests.
Most of these are fast unit tests, but the pipeline also includes a few smoke tests with medium duration.

After a passing \enquote{check} and an approving code review, the pull request (PR) can be scheduled for merging.
This is when the \enquote{gate} pipeline is executed.
It consists of the build and the execution of mostly the same tests that also run in the \enquote{check} pipeline.
The \enquote{gate} ensures that the combination of independently passing changes does not lead to regressions.
After the merge is successfully completed, the \enquote{post} pipeline is triggered that runs up to about \num{1400} tests, most of which are also part of earlier pipelines, and about \num{400} are larger tests that are too slow for earlier pipelines.
Together, these three pipelines execute about up to \num{18000} tests, which is roughly \SI{40}{\percent} of the full test suite.
The remaining tests are scheduled through other pipelines, such as release processes.
All test executions are conducted through Bazel, which
employs a regression test selection based on its dependency graph, resulting in test outcome \texttt{(Cached) Passed} (\cref{fig:ctra_screenshot}) if a test was skipped.

\subsubsection{Ground truth}
To detect flaky tests, the developers of this project already established two rerun-based techniques:
To assess their stability, tests running in the 'check' and 'gate' pipelines are also executed 50 times in separate periodic pipelines that run every weekend.
As hardly any development takes place on weekends, these reruns are conducted on essentially the same software version.
Furthermore, the developers use Bazel's \textit{flaky} attribute~\cite{bazelAttributeFlaky} to rerun certain tests on failure up to three times.
When such tests fail initially but succeed after being rerun, the test result will be \texttt{Flaky} (\cref{fig:ctra_screenshot}).
We use the data coming from the weekly test executions as well as the reruns on failure as the ground truth of our evaluation data set:
We regard a test case as flaky at a time of execution, if it showed different outcomes during the weekend pipeline executions preceding and following this execution, or if flakiness was indicated by a rerun-on-failure following this execution.
Note that although we utilize pre-existing reruns for our evaluation, our approach does neither rely on, nor cause any additional reruns.
Our models could also be trained using other sources of ground truth, for example fixed flaky tests.

\subsubsection{Sampling}%
\label{sec:evaluation_dataset_sampling}

Based on our ground truth criterion, we collected 100 flaky and 100 non-flaky units between November 2021 and May 2022 by means of random sampling.
This process resulted in a balanced data set with \num{200} records.
Note that one unit always refers to one test case at a specific time, meaning that two units can refer to the same test at different times.
This case does in fact occur: the 200 sampled units refer to 182 distinct test cases, with 18 tests being referred to by two units.
However, we do not expect this to create a bias because of two reasons.
First, for most of these 18 tests the observation windows of their two units are several months apart, and second, 16 of these tests even have different labels in their two units, meaning that they were flaky in one unit, but non-flaky in the other.
Sampling non-flaky units posed a special challenge, since most non-flaky tests simply always yield the same verdict (most always pass, few fail for longer periods), meaning that they are trivially distinguishable by their flip rate.
The more difficult, but also more desirable distinction is the one between flaky failures and failures caused by regressions.
When sampling non-flaky tests, we therefore manually searched for tests that did both pass and fail, but were not flaky, i.e., their failures were caused by recent regressions.
In order to reduce the test selection bias that may be introduced by this strategy, we included 15 \enquote*{more typical} non-flaky tests that have a failure rate of either zero or one, meaning that they always passed or always failed.%

\subsubsection{Feature collection (test history)}
For each unit, we collected its test outcomes and durations over the last 3 months or a maximum of \num{10000} test executions.
Since some failures are rerun up to three times to detect flaky tests, this history also contains the outcome \texttt{Flaky}, indicating that one of these reruns passed.
If, on the other hand, all reruns failed, the test is reported as \texttt{Failed}.
Note that although multiple executions of the same test case were conducted in case of a rerun-on-failure, they are still treated as one run with one outcome.
Since we aim to construct a flakiness detection approach that does not rely on reruns, we replace the test outcome \texttt{Flaky} by \texttt{Failed}, simulating that no reruns were conducted after the initial failure.
For the same reason, we also exclude all test executions conducted through the periodic weekend pipelines since their sole purpose is to detect flaky tests.
Furthermore, we filter out all \texttt{(CACHED) PASSED} results, since  no actual test executions took place due to test selection.
In total, we collect slightly above one million test executions.

\subsubsection{Feature collection (code evolution)}%
\label{sec:evaluation_dataset_featureCollection_codeEvolution}
For each unit, we extract code churn information from its parenting Git submodule.
Overall, we collect the following features:
(a)~In the observation windows of \num{3}, \num{14}, and \num{54} days we came across a total of \num{38} different file extensions within the changed files, resulting in $3 \cdot 38 = 114$ features.
(b)~The 9 submodules of the project are encoded as one-hot feature vectors.
(c)~The number of modified files in the current pull request creates one feature, ranging from \num{1} to \num{457} with a median of \num{5}.
(d)~The number of distinct contributors involved in the current pull request is a single feature that ranges from \num{1} to \num{4} with a median of \num{1}.
In total, we extracted $114 + 9 + 1 + 1 = 125$ code churn features.

\subsection{Evaluation Methodology}

\subsubsection{Baseline}
To evaluate the flakiness detection capabilities of our approach, we compare the predictive performance of our models both against each other and against the no-five-failures-in-a-row heuristic described in~\cref{sec:background_flakiness_detection_in_practice}.
We use the latter as a baseline since our approach is meant to be applicable in this very context and also to replace this strategy in case it yields better results.

\subsubsection{Feature combinations}
For RQ1, we evaluate seven different feature sets, each containing one feature created by different ways to process the test result history (entropy + 6 different decay functions applied to the flip rate).
In RQ2 we combine the best performing feature from RQ1 with the two duration-based features, first individually to differentiate their contributions, then all combined, assessing a total of three feature sets.
Symmetrically, in RQ3, we combine the best-performing feature from RQ1 with the code churn information, resulting in one feature set of 126 features.
Since the code churn data consist of 125 individual features, we cannot exhaustively assess all possible combinations.
RQ4 addresses this and evaluates two feature sets:
First, the full feature set, containing 128 features (best feature of RQ1 + two duration-based features + \num{125} code churn features).
Second, a subset created by means of a feature selection based on SHAP values~(\cref{sec:evaluation_methodology_SHAP}), aiming to evaluate which features are most important to the prediction and to reduce the complexity of the model.

\subsubsection{Classifier selection}%
\label{sec:evaluation_classifier_selection}

Since in RQ1 we only consider one feature at a time and want to avoid overfitting, we limit our training to a simple threshold decision, i.e., a decision tree with depth~1.
For all other RQs we train our models based on the following 11 established supervised learning classifiers, using the Scikit-learn library~\cite{scikit-learn}:
Decision Tree, Ada Boost, Naive Bayes, Linear Discriminant Analysis, Quadratic Discriminant Analysis, Random Forest, Multi-layer Perceptron, Gradient Boosting, kNN, Support Vector Machine, and Gaussian Process classifier.
This set is a superset of the classifiers used by the FlakeFlagger study~\cite{alshammari2021flakeflagger}. 
Since accumulated over all feature sets of RQ2 to RQ4, the Gradient Boosting Classifier---which was also used by Machalica et al.~\cite{machalica2019predictive}---yields the best overall performance, we only report the results of this classifier, but make all results available in our replication package\cite{practicalReplicationPackage}.
The Random Forest, which has been used by several existing flakiness detection approaches~\cite{pinto2020what, alshammari2021flakeflagger}, was the second-best performing classifier in our evaluation.

\subsubsection{Training and testing}

To train and evaluate our models, we use cross-validation by means of a stratified five-fold split and report the mean precision, recall, and F1-score reached on the test set across all five folds.

\subsubsection{Estimating the stability of threshold classifications}
To assess the stability of threshold decisions (RQ1), we compute the \textit{relative standard deviation} in the threshold within the five cross-validation folds.
The relative standard deviation, or coefficient of variation, $c_v$ is defined as the ratio of the standard deviation~$\sigma$ and the mean~$\mu$: $c_v = \frac{\sigma}{\mu}$.
We prefer this value over the standard deviation, since the thresholds themselves differ in orders of magnitude depending on the applied decay function, making their standard deviations not directly comparable.
A high relative standard deviation indicates that the optimal threshold strongly dependents on the specific train-test split of the data set and might not generalize well.
A low relative standard deviation on the other hand indicates a more stable performance, independently of the specific split.

\subsubsection{Explaining complex classifiers \& feature selection}%
\label{sec:evaluation_methodology_SHAP}
To assess which features are most important to more complex classifiers and how they contribute to the prediction, we compute SHAP values~\cite{lundberg2017unified} and visualize them in beeswarm plots.
SHAP (SHapley Additive exPlanations) is widely used in the field of artificial intelligence (AI) to explain the behavior that AI models achieve after training, i.e., the effect of each input feature on a specific prediction. 
In general, SHAP assigns to each feature an importance value for a particular prediction.
We apply SHAP to assess the impact of each input feature on the model outcome in terms of the type, i.e., for or against test flakiness, and magnitude of the decisive force.
Hence, the SHAP values determine which features indicate or suppress the decision to test flakiness. 
Since SHAP also identifies the most important features that efficiently describe test flakiness, we additionally use it to reduce the input space by omitting less significant features (feature selection)---resulting in an efficient and representative subset of features that is computationally easier to handle and interpretable for testers.
We calculate the SHAP values using the same cross-validation strategy that we employ for evaluating the predictive performance.

\begin{table*}[]
\caption{Overview of the evaluation results.}
\label{tab:results_overview}
\centering
\begin{tabular}{lllrrr}
\toprule
                                                                       & Feature Set                                                     & Classifier                & Precision                                                     & Recall                                                     & F1-score     \\
\midrule
\textit{Baseline}                                                      & Test Result History                                             & no five failures in a row & \baselinePrecision                                            & \baselineRecall                                            & \baselineFOne \\ \\
\textit{RQ1}                                                           & Unweighted Flip Rate                                            & Threshold                 & \flipRateConstantDtOnePrecisionMeanPercent                    & \flipRateConstantDtOneRecallMeanPercent                    & \flipRateConstantDtOneFOneMeanPercent \\
                                                                       & Weighted Flip Rate, Reciprocal-squared                          & Threshold                 & \flipRateReciprocalSquaredDtOnePrecisionMeanPercent           & \flipRateReciprocalSquaredDtOneRecallMeanPercent           & \flipRateReciprocalSquaredDtOneFOneMeanPercent \\
                                                                       & Entropy                                                         & Threshold                 & \entropyDtOnePrecisionMeanPercent                             & \entropyDtOneRecallMeanPercent                             & \entropyDtOneFOneMeanPercent \\ \\
\textit{RQ2}                                                           & Flip rate rec.-sq. + Mean duration                              & Gradient Boosting         & \flipRateXmeanDurationGbPrecisionMeanPercent                  & \flipRateXmeanDurationGbRecallMeanPercent                  & \flipRateXmeanDurationGbFOneMeanPercent \\
                                                                       & Flip rate rec.-sq. + Mean duration diff                         & Gradient Boosting         & \flipRateXmeanDurationDiffGbPrecisionMeanPercent              & \flipRateXmeanDurationDiffGbRecallMeanPercent              & \flipRateXmeanDurationDiffGbFOneMeanPercent \\
                                                                       & Flip rate rec.-sq. + Mean duration + Mean duration diff         & Gradient Boosting         & \flipRateXmeanDurationXmeanDurationDiffGbPrecisionMeanPercent & \flipRateXmeanDurationXmeanDurationDiffGbRecallMeanPercent & \flipRateXmeanDurationXmeanDurationDiffGbFOneMeanPercent \\ \\
\textit{RQ3}                                                           & Flip rate rec.-sq. + code churn features                        & Gradient Boosting         & \flipRateXcodeChurnGbPrecisionMeanPercent                     & \flipRateXcodeChurnGbRecallMeanPercent                     & \flipRateXcodeChurnGbFOneMeanPercent       \\ \\
\textit{RQ4}                                                                      & Full                                                            & Gradient Boosting         & \fullGbPrecisionMeanPercent                                   & \fullGbRecallMeanPercent                                   & \fullGbFOneMeanPercent \\
                                               & Flip rate rec.-sq + CPP changes last 54 day + no. changed files & Gradient Boosting         & \flipRateXcppFiftyfourXchangedFilesGbPrecisionMeanPercent     & \flipRateXcppFiftyfourXchangedFilesGbRecallMeanPercent     & \flipRateXcppFiftyfourXchangedFilesGbFOneMeanPercent \\
\bottomrule
\end{tabular}
\end{table*}

\subsection{Threats to Validity}

\subsubsection{External validity}

Similar to other industrial studies~\cite{strandberg2020intermittently},
we evaluated our approach on a single large industrial software project.
This may lead to limited generalizability of our results.
Unfortunately, we were unable to apply our approach to existing data sets of flaky tests in open source projects, such as IDoFT~\cite{InternationalDatasetofFlakyTests} or flakiness in Python~\cite{gruber2021empirical}, since these only contain the flaky tests, but not the historic test results.
While we could have conducted artificial test executions of previous versions, we would not have been able to simulate real-world conditions, such as temporarily high CI loads.
Using our evaluation setup instead, we were able to sample test executions that were conducted under the developer-intended setup from a real-world production CI environment.
Overall, more evaluations on larger and more diverse data sets are needed to confirm the effectiveness of our flakiness detection approach.

\subsubsection{Construct validity}

Since our approach targets industry environments in which existing flakiness detection approaches are not applicable, we can only compare the predictive power of our new approach against the current industry practice but not against other published flakiness detection methods.
To do so, comparison studies would need to collect all the information required by those techniques.

\subsubsection{Internal validity}

As a ground truth criterion for our evaluation, we rely on existing test reruns that were triggered periodically (50 reruns), and by test failures (up to 3 reruns).
This may be insufficient to correctly label each test case.
Furthermore, our sampling method of non-flaky tests (\cref{sec:evaluation_dataset_sampling}) might influence the results.
However, since we work on a balanced data set, the threat of having many false negatives is mitigated by the overall sparseness of flaky tests compared to non-flaky ones (practitioners report \SI{1.5}{\percent} to \SI{16}{\percent} of their tests to be flaky~\cite{micco2016flaky}).

\subsection{Baseline: No Five Failures in a Row}

Since the baseline heuristic is created through expert knowledge rather than machine learning, we do not perform cross-validation for assessing its predictive performance.
Instead, we use the entire data set for the evaluation.
The first row in~\cref{tab:results_overview} summarizes its results:
Overall it reached a precision of \baselinePrecision, a recall of \baselineRecall, and a resulting F1-score of \baselineFOne.
The considerably higher recall than precision shows that the baseline is a rather conservative estimate.

\subsection{RQ1: Test Outcomes}%
\label{sec:evaluation_rq1}

\cref{tab:rq1_comparison} holds the evaluation results for different weighting functions applied to the flip rate, as well as for the entropy.
The unweighted flip rate performed very poorly: its F1-score of \flipRateConstantDtOneFOneMeanPercent is below the baseline and its high relative standard deviation in the threshold of \flipRateConstantDtOneThresholdRelativeStdPercent ($\mu = 0.0101 $, $\sigma = 0.00257 $) suggests that this method is rather instable.
Looking at weighted flip rates, we see an improvement
with a trend toward stronger decay functions.
The best performing one was the reciprocal-squared weighting with an F1-score of \flipRateReciprocalSquaredDtOneFOneMeanPercent.
Its threshold also has a far lower relative standard deviation, indicating a stable and data-split independent classification.
The observation that stronger decay functions lead to better performance is also confirmed when looking at the exponential weighting function, which decays slower than the linear one and also performs worse with more instability.
We suspect that stronger decay functions allow the model to correctly label non-flaky tests that have been flaky in the past.
This claim is supported by the fact that stronger decay functions almost only improve the precision, but not the recall (fewer false positives).
The exponentially weighted moving average used by Kowalczyk et al.~\cite{kowalczyk2020modeling} delivered the second-best results, with an F1-score of \flipRateEwmaDtOneFOneMeanPercent.
Surprisingly, the unweighted entropy-based classification was able to outperform the unweighted flip rate and, with an F1-score of \entropyDtOneFOneMeanPercent, achieved the fourth-best result among models using only the test result history.

\summary{RQ1: Only test outcomes}{
    Unexpectedly, the unweighted flip rate model performs worse than the baseline.
    The predictive performance improves with stronger decay functions, however, with an F1-score of \flipRateReciprocalSquaredDtOneFOneMeanPercent, even the best-performing reciprocal-squared weighting only slightly exceeds the baseline.
}

\subsection{RQ2: Test Outcomes + Test Durations}

Here we combine the best-performing feature of RQ1 (the reciprocal-squared weighted flip rate) with one or two of the test duration features (the mean test duration, and/or the difference between the average passing and failing duration).
The RQ2-group of \cref{tab:results_overview} hold the evaluation results.
Overall, adding both duration features improves the predictive performance of our model significantly and achieves an average F1-score of \flipRateXmeanDurationXmeanDurationDiffGbFOneMeanPercent.
While the models in RQ1 reach a substantially higher precision than recall, the RQ2 classifications are much more balanced.
To examine which of the two test duration features contributes most to the improvement, we combine each of them individually with the flip rate.
Surprisingly, their impact is very different, with the mean duration yielding only minor improvements while the
mean duration diff
substantially enhances the performance.

\begin{figure}
    \centering
    \includegraphics[width=\linewidth]{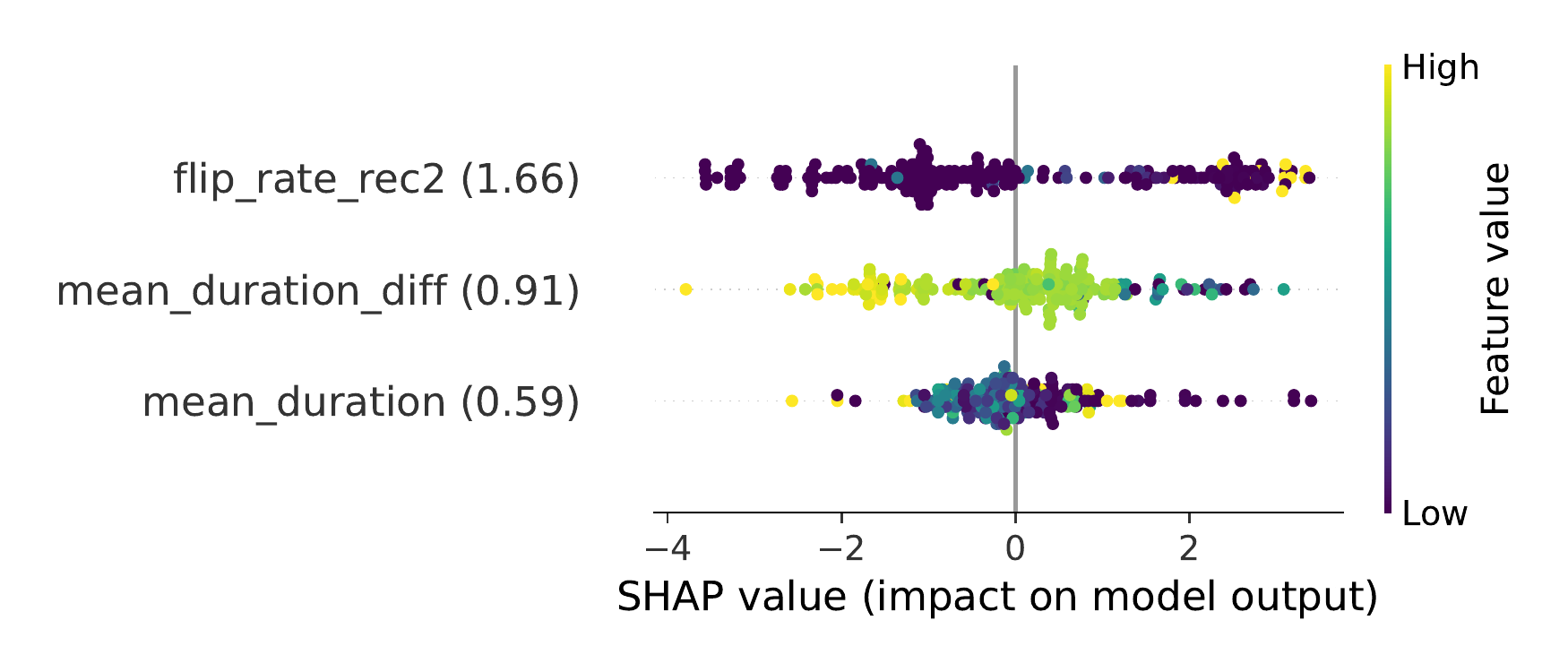}
    \caption{Test outcomes and durations, beeswarm plot of SHAP values (mean SHAP values per feature).}
    \label{fig:shap_rq2}
\end{figure}

\cref{fig:shap_rq2} depicts the SHAP values for the model trained on all three features.
The features are displayed in descending order by their mean absolute SHAP values, which is a measure of the broad average impact of the feature.
Each row concerns one feature, and each dot within a row represents the respective feature value of one data point.
Bright colors indicate high feature values, dark colors indicate low feature values.
Dots on the left (negative x-values) suggest that this feature value contributed to a negative prediction, i.e., non-flaky.
Dots on the right (positive x-values) suggest that the feature value contributed to a positive prediction, i.e., flaky.
The distance from the center (x = 0) is proportional to the strength of the influence on the prediction.
The height of a dot within a row (y-value) does not make any statement about the SHAP value itself, but indicates the density in this area.
As we can see, the weighted flip rate had the most impact on the prediction and high flip rates strongly lead to a positive prediction, meaning that the test case is suggested to be flaky.
The second most important feature---as we already found through adding features individually---is the
mean duration diff.
In this case, the beeswarm plot suggests a negative correlation.
Following our intuition discussed in~\cref{sec:test_duration}, this may be caused by timeouts, leading to long-lasting flaky failures in comparison to failures caused by regressions.
Apart from advancing our flakiness prediction, this insight also contributes to our understanding of the problem and additionally demonstrates the impact of flakiness in the form of delayed test feedback and additional computational effort.
The least important feature of this model was the mean duration.
Against our expectation, longer-lasting tests did not show a tendency towards flakiness.

\summary{RQ2: Test outcomes and durations}{
    Consulting the average test duration in addition to the test outcomes does not substantially improve the predictive performance of our approach, however, adding the average duration difference between passing and failing test executions does.
}

\begin{table}[]
    \caption{Predictive performance of flip rate and entropy.}
    \label{tab:rq1_comparison}
    \centering
    \begin{tabular}{lrrrr}
    \toprule
        Metric / Decay Func.        & Precision                                      & Recall                                      & F1                                 & $c_v$(threshold)                           \\
    \midrule
        Constant (unweighted) & \flipRateConstantDtOnePrecisionMeanPercent          & \flipRateConstantDtOneRecallMeanPercent          & \flipRateConstantDtOneFOneMeanPercent          & \flipRateConstantDtOneThresholdRelativeStdPercent          \\
        Linear                & \flipRateLinearDtOnePrecisionMeanPercent            & \flipRateLinearDtOneRecallMeanPercent            & \flipRateLinearDtOneFOneMeanPercent            & \flipRateLinearDtOneThresholdRelativeStdPercent            \\
        Exponential           & \flipRateExponentialDtOnePrecisionMeanPercent       & \flipRateExponentialDtOneRecallMeanPercent       & \flipRateExponentialDtOneFOneMeanPercent       & \flipRateExponentialDtOneThresholdRelativeStdPercent       \\
        Reciprocal            & \flipRateReciprocalDtOnePrecisionMeanPercent        & \flipRateReciprocalDtOneRecallMeanPercent        & \flipRateReciprocalDtOneFOneMeanPercent        & \flipRateReciprocalDtOneThresholdRelativeStdPercent        \\
        Reciprocal Squared    & \flipRateReciprocalSquaredDtOnePrecisionMeanPercent & \flipRateReciprocalSquaredDtOneRecallMeanPercent & \flipRateReciprocalSquaredDtOneFOneMeanPercent & \flipRateReciprocalSquaredDtOneThresholdRelativeStdPercent \\
        EWMA & \flipRateEwmaDtOnePrecisionMeanPercent        & \flipRateEwmaDtOneRecallMeanPercent        & \flipRateEwmaDtOneFOneMeanPercent        & \flipRateEwmaDtOneThresholdRelativeStdPercent        \\
    \\
        Entropy (unweighted) & \entropyDtOnePrecisionMeanPercent & \entropyDtOneRecallMeanPercent & \entropyDtOneFOneMeanPercent & \entropyDtOneThresholdRelativeStdPercent \\
    \bottomrule
    \end{tabular}
\end{table}

\subsection{RQ3: Test Outcomes + Code Churn Information}

\begin{figure}
    \centering
    \includegraphics[width=\linewidth]{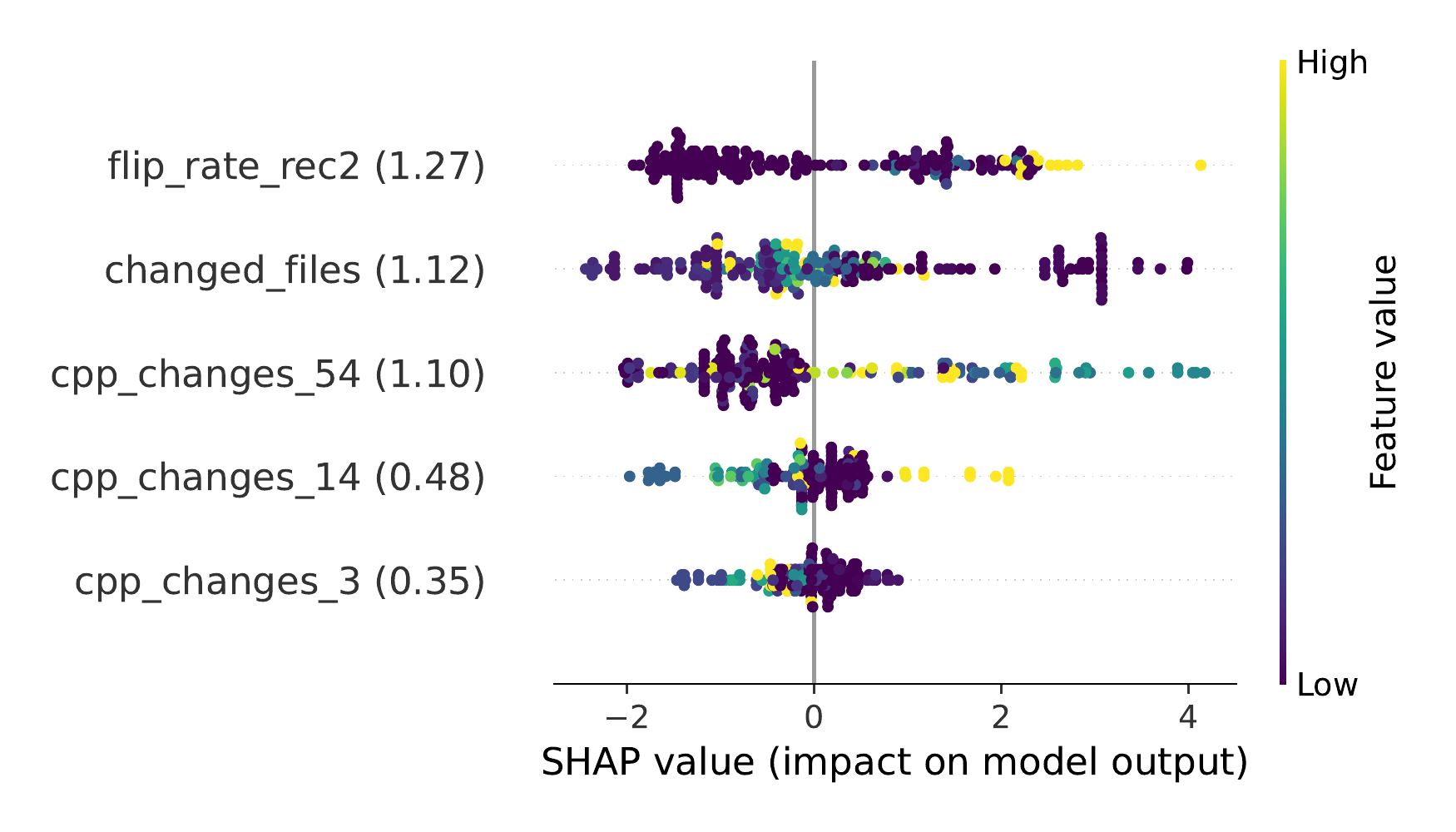}
    \caption{Flip rate + code churn, SHAP values, top 5 features.}
    \label{fig:shap_rq3}
\end{figure}

\cref{tab:results_overview} also shows the predictive performance when combining the weighted flip rate with code churn features.
With an average F1-score of \flipRateXcodeChurnGbFOneMeanPercent there is a significant improvement over all previously evaluated feature sets.
This is especially noteworthy since Bazel's regression test selection reduces the number of reruns conducted on effectively unchanged code.
\cref{fig:shap_rq3} depicts the SHAP values of the 5 most important features.
As in the previous RQ, the most important feature to the prediction is the flip rate, this time followed by
the number of files changed in the current pull request
and
the number of changes to C++ files in the last 54 days.
As expected, a low number of changed files is a positive indicator for flakiness, as failures occurring after such changes are less likely to be caused by regressions.

Against our intuition, however, a large number of long-term source code changes (in this case C++) leads to a positive prediction (flaky).
A possible explanation for this phenomenon could be that by measuring the number of long-term C++ file changes, we indirectly captured submodules that make strong use of C++, which offers more possible sources for flakiness compared to other languages, such as uninitialized variables or differences between compilers~\cite{gruber2022survey}.
Looking at short-term source code changes (\texttt{cpp\_changes\_3}), we see the opposite and also expected picture with high feature values leading to a negative prediction (non-flaky), and low numbers of source code changes leading to a positive prediction (flaky).
However, this feature is less important to the prediction (lower mean absolute SHAP values), which is why we do not discuss it further, like all features beyond the top three.
It is worth noting that none of the project name features appeared among the 5 most important features, suggesting that our data set is not biased due to flakiness existing only in a specific submodule.

\summary{RQ3: Test outcomes and code churn}{
    Adding information about code evolution to the flip rate significantly improves flakiness detection.
    The most important features are the number of changed files in the current pull request, as well as the number of long-term source file changes.
}

\subsection{RQ4: Full Feature Set + Feature Selection}

\begin{figure}
    \centering
    \includegraphics[width=\linewidth]{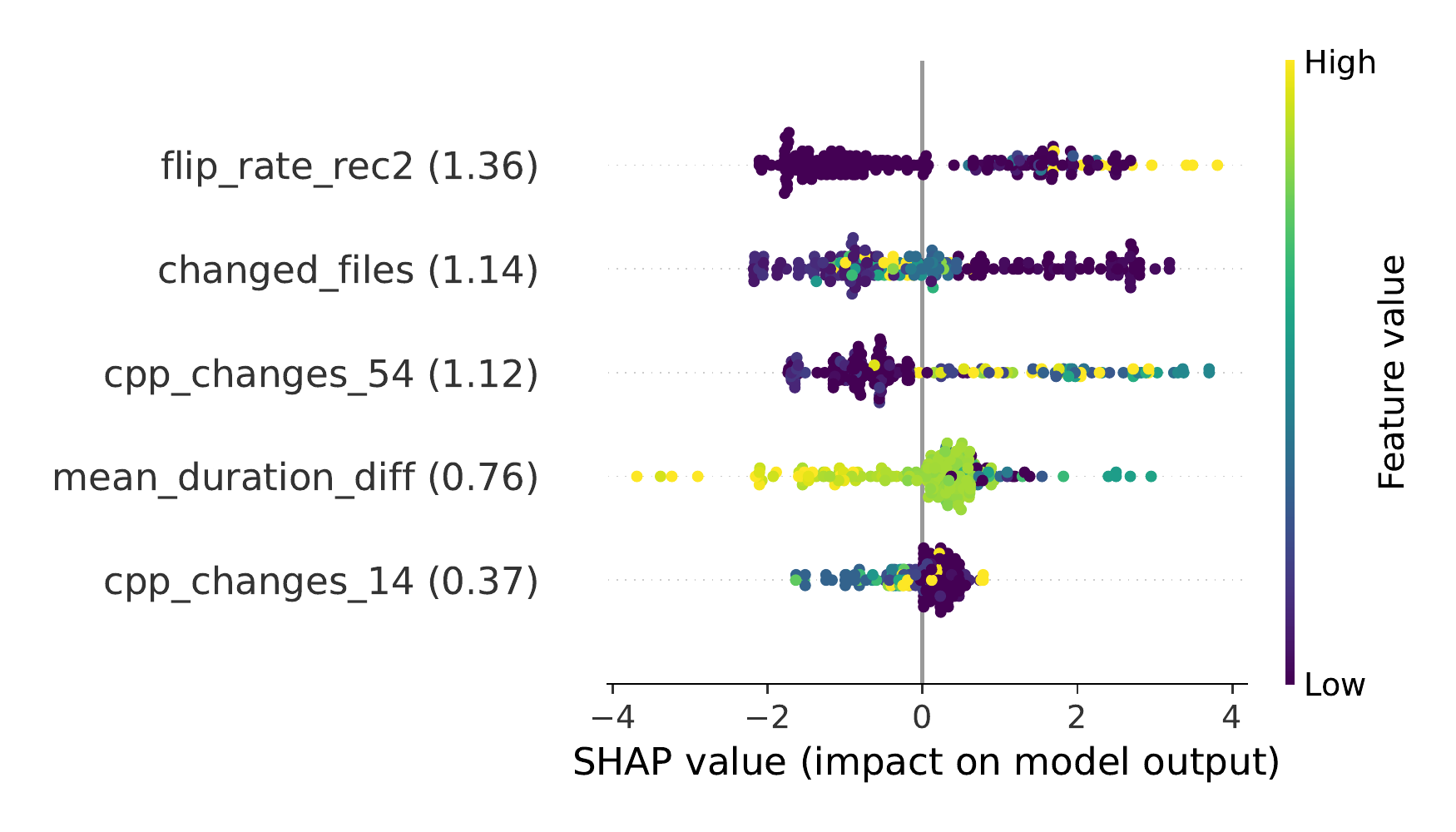}
    \caption{Full feature set, SHAP values, top 5 features.}
    \label{fig:shap_full}
\end{figure}

The second last row of~\cref{tab:results_overview} shows the average cross-evaluated performance of the full feature set including the weighted flip rate, the mean duration, the mean difference between passing and failing durations, and the code churn features.
Compared to the previous feature set in RQ3 (without duration features), we see a similar performance, suggesting that the tests' durations do not play an important role in the prediction of flakiness, emphasizing again the relevance of the code churn features (comparison to RQ1 and RQ2).
This is confirmed by the SHAP values depicted in~\cref{fig:shap_full} where we again see the weighted flip rate, the number of changed files in the current pull request, and the number of C++ changes in the last 54 days to be the most important features.

As described in~\cref{sec:evaluation_methodology_SHAP}, we also use the SHAP values to apply feature selection, aiming to simplify the model and avoid overfitting.
Looking at the mean absolute SHAP values, we can observe a significant drop after the third most important feature, which is why we decided to consider only the top three features for our feature subset.
We used these to train a new model, whose predictive performance is depicted in the last row of \cref{tab:results_overview}:
With an F1-score of \flipRateXcppFiftyfourXchangedFilesGbFOneMeanPercent it slightly outperforms all previous models and therefore underlines that the strong positive influence of the code churn features did not stem from overfitting on \num{125} different features.

\summary{RQ4: Full feature set + feature selection}{
    Using feature selection on the full feature set, we achieve an F1-score of \flipRateXcppFiftyfourXchangedFilesGbFOneMeanPercent using only three features, highlighting the flakiness prediction capabilities of the weighted flip rate and the code churn features.
}

\section{Related work}

The Background~\cref{sec:background} already discussed published approaches to flakiness detection and compared them to our approach.
This section therefore focuses on the role of flakiness detection in the context of different root causes of flakiness and the distinction to Machalica et al.~\cite{machalica2019predictive}.

Luo et al.~\cite{luo2014empirical} conducted the first empirical study on flaky tests.
They found asynchronous waiting and concurrency to be the most common root causes---which other studies confirmed \cite{eck2019understanding, ahmad2021empirical, habchi2022qualitative, gruber2022survey, parry2022surveying}---and recommended that detection techniques should focus on these.
We take this observation into account by considering the difference between the tests' mean passing and failing durations and find that flaky failures tend to take longer than failures caused by regressions, which we suspect to be the result of waiting for asynchronous processes.

Order-dependency (OD) is a special cause of flakiness that has received much attention from researchers proposing efficient techniques to detect it~\cite{gambi2018practical, lam2019idflakies, wei2021probabilistic, wei2022preempting, li2022evolution}.
Due to their special nature (good reproducibility since tests behave deterministic when run in the same order), it makes sense to use dedicated methods to search for OD flaky tests, however, existing techniques also rely on instrumentation and reruns.

To extract features from code evolution, we were inspired by Machalica et al.\cite{machalica2019predictive}, who used this information to build a predictive test selection strategy.
They also address the issue of test flakiness, however, their goal is to minimize its impact on regression test selection.
Our goal, on the other hand, is to detect if a test is flaky without using reruns.

\section{Conclusions}

While attempting to deploy flakiness detection approaches in an industrial context, we found a mismatch between the strong technical demands of methods proposed by researchers and the available information and possibilities in practice.
To close this gap and make flakiness detection available to a wide range of developers, we proposed a generic, predictive method to classify tests as flaky based only on the most widely available information that can be derived from continuous integration and version control systems.
We evaluated this technique on a large industrial software project.
Our best model achieved an F1-score of \flipRateXcppFiftyfourXchangedFilesGbFOneMeanPercent using only the weighted flip rate, the number of changed files in the current pull request, and the number of changes to C++ files in the last 54 days.
This indicates that our approach can be effectively and practically applied to detect flaky tests.

In order to support future research and applications, our features, models, and data are available~\cite{practicalReplicationPackage}.


\begin{thebibliography}{10}
\providecommand{\url}[1]{#1}
\csname url@samestyle\endcsname
\providecommand{\newblock}{\relax}
\providecommand{\bibinfo}[2]{#2}
\providecommand{\BIBentrySTDinterwordspacing}{\spaceskip=0pt\relax}
\providecommand{\BIBentryALTinterwordstretchfactor}{4}
\providecommand{\BIBentryALTinterwordspacing}{\spaceskip=\fontdimen2\font plus
\BIBentryALTinterwordstretchfactor\fontdimen3\font minus
  \fontdimen4\font\relax}
\providecommand{\BIBforeignlanguage}[2]{{%
\expandafter\ifx\csname l@#1\endcsname\relax
\typeout{** WARNING: IEEEtran.bst: No hyphenation pattern has been}%
\typeout{** loaded for the language `#1'. Using the pattern for}%
\typeout{** the default language instead.}%
\else
\language=\csname l@#1\endcsname
\fi
#2}}
\providecommand{\BIBdecl}{\relax}
\BIBdecl

\bibitem{parry2021survey}
O.~Parry, G.~M. Kapfhammer, M.~Hilton, and P.~McMinn, ``A survey of flaky
  tests,'' \emph{{ACM} Transactions on Software Engineering and Methodology},
  pp. 1--74, 2021.

\bibitem{fowler2011eradicating}
\BIBentryALTinterwordspacing
M.~Fowler, ``Eradicating non-determinism in tests,'' 2011. [Online]. Available:
  \url{https://martinfowler.com/articles/nonDeterminism.html}
\BIBentrySTDinterwordspacing

\bibitem{micco2016flaky}
\BIBentryALTinterwordspacing
J.~Micco, ``Flaky tests at {Google} and how we mitigate them,'' 2016. [Online].
  Available:
  \url{https://testing.googleblog.com/2016/05/flaky-tests-at-google-and-how-we.html}
\BIBentrySTDinterwordspacing

\bibitem{oezal2021how}
\BIBentryALTinterwordspacing
S.~Özal, ``How to deal with flaky tests,'' 2021. [Online]. Available:
  \url{https://thenewstack.io/how-to-deal-with-flaky-tests/}
\BIBentrySTDinterwordspacing

\bibitem{champier2017hunting}
\BIBentryALTinterwordspacing
C.~Champier, ``Hunting flaky tests at {Doctolib},'' 2017. [Online]. Available:
  \url{https://medium.com/doctolib/hunting-flaky-tests-at-doctolib-6686944ab426}
\BIBentrySTDinterwordspacing

\bibitem{palmer2019test}
\BIBentryALTinterwordspacing
J.~Palmer, ``Test flakiness – methods for identifying and dealing with flaky
  tests,'' 2019. [Online]. Available:
  \url{https://engineering.atspotify.com/2019/11/18/test-flakiness-methods-for-identifying-and-dealing-with-flaky-tests/}
\BIBentrySTDinterwordspacing

\bibitem{raine2020reducing}
\BIBentryALTinterwordspacing
J.~Raine, ``Reducing flaky builds by 18x,'' 2020. [Online]. Available:
  \url{https://github.blog/2020-12-16-reducing-flaky-builds-by-18x/}
\BIBentrySTDinterwordspacing

\bibitem{lam2020study}
W.~Lam, K.~Mu{\c{s}}lu, H.~Sajnani, and S.~Thummalapenta, ``A study on the
  lifecycle of flaky tests,'' in \emph{International Conference on Software
  Engineering~(ICSE)}, 2020, pp. 1471--1482.

\bibitem{lam2019root}
W.~Lam, P.~Godefroid, S.~Nath, A.~Santhiar, and S.~Thummalapenta, ``Root
  causing flaky tests in a large-scale industrial setting,'' in
  \emph{International Symposium on Software Testing and Analysis~(ISSTA)},
  2019, pp. 101--111.

\bibitem{terragni2020container}
V.~Terragni, P.~Salza, and F.~Ferrucci, ``A container-based infrastructure for
  fuzzy-driven root causing of flaky tests,'' in \emph{International Conference
  on Software Engineering: New Ideas and Emerging Results~(ICSE-NIER)}, 2020,
  pp. 69--72.

\bibitem{celalziftci2020de}
D.~C. Celal~Ziftci, ``{De-Flake} your tests: Automatically locating root causes
  of flaky tests in code at {Google},'' in \emph{International Conference on
  Software Maintenance and Evolution~(ICSME)}, 2020, pp. 736--745.

\bibitem{akli2022predicting}
A.~Akli, G.~Haben, S.~Habchi, M.~Papadakis, and Y.~L. Traon, ``Predicting flaky
  tests categories using few-shot learning,'' \emph{CoRR}, 2022.

\bibitem{shi2019ifixflakies}
A.~Shi, W.~Lam, R.~Oei, T.~Xie, and D.~Marinov, ``{iFixFlakies}: {A} framework
  for automatically fixing order-dependent flaky tests,'' in \emph{Joint
  Meeting of the European Software Engineering Conference and the Symposium on
  the Foundations of Software Engineering~(ESEC/FSE)}, 2019, pp. 545--555.

\bibitem{wang2022ipflakies}
W.~L. Ruixin~Wang, Yang~Chen, ``{iPFlakies}: A framework for detecting and
  fixing python order-dependent flaky tests,'' in \emph{International
  Conference on Software Engineering: Companion Proceedings~(ICSE Companion)},
  2022, pp. 120--124.

\bibitem{alshammari2021flakeflagger}
A.~Alshammari, C.~Morris, M.~Hilton, and J.~Bell, ``{FlakeFlagger}:
  {Predicting} flakiness without rerunning tests,'' in \emph{International
  Conference on Software Engineering~(ICSE)}, 2021, pp. 1572--1584.

\bibitem{parry2022evaluating}
O.~Parry, G.~M. Kapfhammer, M.~Hilton, and P.~McMinn, ``Evaluating features for
  machine learning detection of order-and non-order-dependent flaky tests,'' in
  \emph{International Conference on Software Testing, Verification and
  Validation~(ICST)}, 2022, pp. 93--104.

\bibitem{shi2016detecting}
A.~Shi, A.~Gyori, O.~Legunsen, and D.~Marinov, ``Detecting assumptions on
  deterministic implementations of non-deterministic specifications,'' in
  \emph{International Conference on Software Testing, Verification and
  Validation~(ICST)}, 2016, pp. 80--90.

\bibitem{bell2018deflaker}
J.~Bell, O.~Legunsen, M.~Hilton, L.~Eloussi, T.~Yung, and D.~Marinov,
  ``{DeFlaker}: {Automatically} detecting flaky tests,'' in \emph{International
  Conference on Software Engineering~(ICSE)}, 2018, pp. 433--444.

\bibitem{dutta2020detecting}
S.~Dutta, A.~Shi, R.~Choudhary, Z.~Zhang, A.~Jain, and S.~Misailovic,
  ``Detecting flaky tests in probabilistic and machine learning applications,''
  in \emph{International Symposium on Software Testing and Analysis~(ISSTA)},
  2020, pp. 211--224.

\bibitem{silva2020shake}
D.~Silva, L.~Teixeira, and M.~d’Amorim, ``Shake it! {Detecting} flaky tests
  caused by concurrency with {Shaker},'' in \emph{International Conference on
  Software Maintenance and Evolution~(ICSME)}, 2020, pp. 301--311.

\bibitem{lam2019idflakies}
W.~Lam, R.~Oei, A.~Shi, D.~Marinov, and T.~Xie, ``{iDFlakies}: A framework for
  detecting and partially classifying flaky tests,'' in \emph{International
  Conference on Software Testing, Verification and Validation~(ICST)}, 2019,
  pp. 312--322.

\bibitem{kowalczyk2020modeling}
E.~Kowalczyk, K.~Nair, Z.~Gao, L.~Silberstein, T.~Long, and A.~Memon,
  ``Modeling and ranking flaky tests at {Apple},'' in \emph{International
  Conference on Software Engineering: Software Engineering in Practice
  Track~(ICSE-SEIP)}, 2020, pp. 110--119.

\bibitem{machalica2020how}
\BIBentryALTinterwordspacing
M.~Machalica, W.~Chmiel, S.~Swierc, and R.~Sakevych, ``How do you test your
  tests?'' 2020. [Online]. Available:
  \url{https://engineering.fb.com/2020/12/10/developer-tools/probabilistic-flakiness/}
\BIBentrySTDinterwordspacing

\bibitem{king2018towards}
T.~M. King, D.~Santiago, J.~Phillips, and P.~J. Clarke, ``Towards a {Bayesian}
  network model for predicting flaky automated tests,'' in \emph{International
  Conference on Software Quality, Reliability and Security
  Companion~({QRS-C})}, 2018, pp. 100--107.

\bibitem{fatima2022flakify}
S.~Fatima, T.~A. Ghaleb, and L.~Briand, ``Flakify: A black-box, language
  model-based predictor for flaky tests,'' \emph{{IEEE} Transactions on
  Software Engineering}, pp. 1--17, 2022.

\bibitem{qin2022peeler}
Y.~Qin, S.~Wang, K.~Liu, B.~Lin, H.~Wu, L.~Li, X.~Mao, and T.~F. D.~A.
  Bissyande, ``Peeler: Learning to effectively predict flakiness without
  running tests,'' in \emph{International Conference on Software Maintenance
  and Evolution~(ICSME)}, 2022, pp. 1--12.

\bibitem{pinto2020what}
G.~Pinto, B.~Miranda, S.~Dissanayake, M.~d'Amorim, C.~Treude, and A.~Bertolino,
  ``What is the vocabulary of flaky tests?'' in \emph{International Conference
  on Mining Software Repositories~(MSR)}, 2020, pp. 492--502.

\bibitem{verdecchia2021know}
R.~Verdecchia, E.~Cruciani, B.~Miranda, and A.~Bertolino, ``Know you neighbor:
  Fast static prediction of test flakiness,'' \emph{IEEE Access}, pp.
  76\,119--76\,134, 2021.

\bibitem{gruber2022survey}
M.~Gruber and G.~Fraser, ``A survey on how test flakiness affects developers
  and what support they need to address it,'' in \emph{International Conference
  on Software Testing, Verification and Validation~(ICST)}, 2022, pp. 82--92.

\bibitem{elbaum2014techniques}
S.~Elbaum, G.~Rothermel, and J.~Penix, ``Techniques for improving regression
  testing in continuous integration development environments,'' in
  \emph{International Symposium on Foundations of Software Engineering~(FSE)},
  2014, pp. 235--245.

\bibitem{micco2017state}
\BIBentryALTinterwordspacing
J.~Micco, ``The state of continuous integration testing @{Google},'' 2017.
  [Online]. Available:
  \url{https://storage.googleapis.com/pub-tools-public-publication-data/pdf/45880.pdf}
\BIBentrySTDinterwordspacing

\bibitem{mavenRerunPlugin}
\BIBentryALTinterwordspacing
``Apache {Maven}: Rerun failing tests.'' [Online]. Available:
  \url{https://maven.apache.org/surefire/maven-surefire-plugin/examples/rerun-failing-tests.html}
\BIBentrySTDinterwordspacing

\bibitem{pythonRerunPlugin}
\BIBentryALTinterwordspacing
``Flaky - a plugin for nose or pytest that automatically reruns flaky tests.''
  [Online]. Available: \url{https://pypi.org/project/flaky/}
\BIBentrySTDinterwordspacing

\bibitem{gruber2021empirical}
M.~Gruber, S.~Lukasczyk, F.~Kroi{\ss}, and G.~Fraser, ``An empirical study of
  flaky tests in {Python},'' in \emph{International Conference on Software
  Testing, Verification and Validation~(ICST)}, 2021, pp. 148--158.

\bibitem{listfield2017where}
\BIBentryALTinterwordspacing
J.~Listfield, ``Where do our flaky tests come from?'' 2017. [Online].
  Available:
  \url{https://testing.googleblog.com/2017/04/where-do-our-flaky-tests-come-from.html}
\BIBentrySTDinterwordspacing

\bibitem{bazelAttributeFlaky}
\BIBentryALTinterwordspacing
``Bazel: Attribute flaky.'' [Online]. Available:
  \url{https://bazel.build/reference/be/common-definitions}
\BIBentrySTDinterwordspacing

\bibitem{strandberg2020intermittently}
P.~E. Strandberg, T.~J. Ostrand, E.~J. Weyuker, W.~Afzal, and D.~Sundmark,
  ``Intermittently failing tests in the embedded systems domain,'' in
  \emph{International Symposium on Software Testing and Analysis~(ISSTA)},
  2020, pp. 337--348.

\bibitem{liviu2019machine}
\BIBentryALTinterwordspacing
S.~Liviu, ``A machine learning solution for detecting and mitigating flaky
  tests,'' 2019. [Online]. Available:
  \url{https://medium.com/fitbit-tech-blog/a-machine-learning-solution-for-detecting-and-mitigating-flaky-tests-c5626ca7e853}
\BIBentrySTDinterwordspacing

\bibitem{champier2019continuous}
\BIBentryALTinterwordspacing
C.~Champier, ``Continuous improvement of our flaky tests hunting process,''
  2019. [Online]. Available:
  \url{https://medium.com/doctolib/continuous-improvement-of-our-flaky-hunting-process-189528a1f540}
\BIBentrySTDinterwordspacing

\bibitem{gradle2020best}
\BIBentryALTinterwordspacing
Gradle, ``Best tool for managing flaky tests,'' 2020. [Online]. Available:
  \url{https://www.youtube.com/watch?v=C5_jgsRANuI}
\BIBentrySTDinterwordspacing

\bibitem{launchableFlakyTestInsights}
\BIBentryALTinterwordspacing
Launchable, ``Flaky test insights.'' [Online]. Available:
  \url{https://docs.launchableinc.com/features/insights/flaky-tests}
\BIBentrySTDinterwordspacing

\bibitem{rudolphbuildpulse}
\BIBentryALTinterwordspacing
J.~Rudolph, ``{BuildPulse}.'' [Online]. Available: \url{https://buildpulse.io/}
\BIBentrySTDinterwordspacing

\bibitem{JUnitXMLSchema}
\BIBentryALTinterwordspacing
``{JUnit-XML-Schema}.'' [Online]. Available:
  \url{https://github.com/windyroad/JUnit-Schema/blob/master/JUnit.xsd}
\BIBentrySTDinterwordspacing

\bibitem{pytestJUnitXML}
\BIBentryALTinterwordspacing
``Pytest: Creating junitxml format files.'' [Online]. Available:
  \url{https://docs.pytest.org/en/7.1.x/how-to/output.html#creating-junitxml-format-files}
\BIBentrySTDinterwordspacing

\bibitem{luo2014empirical}
Q.~Luo, F.~Hariri, L.~Eloussi, and D.~Marinov, ``An empirical analysis of flaky
  tests,'' in \emph{International Symposium on Foundations of Software
  Engineering~(FSE)}, 2014, pp. 643--653.

\bibitem{eck2019understanding}
M.~Eck, F.~Palomba, M.~Castelluccio, and A.~Bacchelli, ``Understanding flaky
  tests: The developer's perspective,'' in \emph{Joint Meeting of the European
  Software Engineering Conference and the Symposium on the Foundations of
  Software Engineering~(ESEC/FSE)}, 2019, pp. 830--840.

\bibitem{parry2022surveying}
O.~Parry, G.~M. Kapfhammer, M.~Hilton, and P.~McMinn, ``Surveying the developer
  experience of flaky tests,'' in \emph{International Conference on Software
  Engineering: Software Engineering in Practice Track~(ICSE-SEIP)}, 2022, pp.
  253--262.

\bibitem{machalica2019predictive}
M.~Machalica, A.~Samylkin, M.~Porth, and S.~Chandra, ``Predictive test
  selection,'' in \emph{International Conference on Software Engineering:
  Software Engineering in Practice Track~(ICSE-SEIP)}, 2019, pp. 91--100.

\bibitem{SOSurvey2022VCS}
\BIBentryALTinterwordspacing
``{Stack Overflow} developer survey: Version control systems,'' 2022. [Online].
  Available:
  \url{https://survey.stackoverflow.co/2022/#version-control-version-control-system-prof}
\BIBentrySTDinterwordspacing

\bibitem{adanial_cloc}
\BIBentryALTinterwordspacing
A.~Danial, ``cloc: v1.92,'' 2021. [Online]. Available:
  \url{https://doi.org/10.5281/zenodo.5760077}
\BIBentrySTDinterwordspacing

\bibitem{bazel}
\BIBentryALTinterwordspacing
``Bazel build.'' [Online]. Available: \url{https://bazel.build/}
\BIBentrySTDinterwordspacing

\bibitem{scikit-learn}
F.~Pedregosa, G.~Varoquaux, A.~Gramfort, V.~Michel, B.~Thirion, O.~Grisel,
  M.~Blondel, P.~Prettenhofer, R.~Weiss, V.~Dubourg, J.~Vanderplas, A.~Passos,
  D.~Cournapeau, M.~Brucher, M.~Perrot, and E.~Duchesnay, ``Scikit-learn:
  Machine learning in {P}ython,'' \emph{J. Machine Learning Research}, pp.
  2825--2830, 2011.

\bibitem{practicalReplicationPackage}
\BIBentryALTinterwordspacing
M.~Gruber, M.~Heine, N.~Oster, M.~Philippsen, and G.~Fraser, ``Practical flaky
  test prediction using common code evolution and test history data
  [replication package].'' [Online]. Available:
  \url{https://doi.org/10.6084/m9.figshare.21363075}
\BIBentrySTDinterwordspacing

\bibitem{lundberg2017unified}
S.~M. Lundberg and S.-I. Lee, ``A unified approach to interpreting model
  predictions,'' in \emph{Advances in Neural Information Processing Systems},
  2017, pp. 4765--4774.

\bibitem{InternationalDatasetofFlakyTests}
\BIBentryALTinterwordspacing
W.~Lam, ``International dataset of flaky tests ({IDoFT}),'' 2020. [Online].
  Available: \url{http://mir.cs.illinois.edu/flakytests}
\BIBentrySTDinterwordspacing

\bibitem{ahmad2021empirical}
A.~Ahmad, O.~Leifler, and K.~Sandahl, ``Empirical analysis of practitioners'
  perceptions of test flakiness factors,'' \emph{Journal of Software Testing,
  Verification and Reliability}, p. e1791, 2021.

\bibitem{habchi2022qualitative}
S.~Habchi, G.~Haben, M.~Papadakis, M.~Cordy, and Y.~L. Traon, ``A qualitative
  study on the sources, impacts, and mitigation strategies of flaky tests,'' in
  \emph{International Conference on Software Testing, Verification and
  Validation~(ICST)}, 2022, pp. 244--255.

\bibitem{gambi2018practical}
A.~Gambi, J.~Bell, and A.~Zeller, ``Practical test dependency detection,'' in
  \emph{International Conference on Software Testing, Verification and
  Validation~(ICST)}, 2018, pp. 1--11.

\bibitem{wei2021probabilistic}
A.~Wei, P.~Yi, T.~Xie, D.~Marinov, and W.~Lam, ``Probabilistic and systematic
  coverage of consecutive test-method pairs for detecting order-dependent flaky
  tests,'' in \emph{International Conference on Tools and Algorithms for
  Construction and Analysis of Systems~(TACAS)}, 2021, pp. 270--287.

\bibitem{wei2022preempting}
A.~Wei, P.~Yi, Z.~Li, T.~Xie, D.~Marinov, and W.~Lam, ``Preempting flaky tests
  via non-idempotent-outcome tests,'' in \emph{International Conference on
  Software Engineering~(ICSE)}, 2022, pp. 1730--1742.

\bibitem{li2022evolution}
C.~Li and A.~Shi, ``Evolution-aware detection of order-dependent flaky tests,''
  in \emph{International Symposium on Software Testing and Analysis~(ISSTA)},
  2022, p. 114–125.

\end{thebibliography}
\balance

\noindent
All online resources accessed on Friday, 31st of October 2022.

\end{document}